# Charge collection efficiency mapping of interdigitated 4H-SiC detectors

E. Vittone[1*], N. Skukan[2], Ž. Pastuović[2], P. Olivero[2,1], M. Jakšić[2]

[1]Experimental Physics Dept./NIS excellence centre, University of Torino and INFN sez. Torino, via P. Giuria 1, 10125 Torino- Italy

[2]Department of Experimental Physics, Ruder Boskovic Institute, P.O. Box 180, 10002 Zagreb, Croatia

**Abstract**

The Ion Beam Induced Charge Collection (IBIC) technique was used to map the charge collection efficiency (CCE) of a 4H-SiC photodetector with coplanar interdigitated Schottky barrier electrodes and a common ohmic contact on the back side.

IBIC maps were obtained using focused proton beams with energies of 0.9 MeV and 1.5 MeV, at different bias voltages and different sensitive electrode configurations (charge collection at the top Schottky or at the back Ohmic contact).

These different experimental conditions have been modeled using a two dimensional finite element code to solve the adjoint carrier continuity equations and the results obtained have been compared with experimental results. The excellent consistency between the simulated and experimental CCE maps allows an exhaustive interpretation of the charge collection mechanisms occurring in pixellated or strip detectors.

**Classification codes and keywords**

**PACS:,** 72.20.Jv Charge carriers: generation, recombination, lifetime and trapping, 41.75.Ak Positive-ion beams, Ion Beam Microscopy

[*] Corresponding author: Ettore Vittone vittone@to.infn.it

Keywords: **IBIC, silicon carbide detector, charge collection efficiency.**

*Introduction*

In previous papers [1][2][3] we presented the mathematical formalism adopted to interpret IBIC experiment on the basis of a solid and rigorous theory founded on the basic laws of electrostatics. The algorithm is based on the solution of the adjoint continuity equations for holes and electrons and allows charge collection efficiency maps to be obtained with reduced computational efforts.

In order to validate this theoretical approach, several benchmark experiments were recently carried out at the Ruđer Bošković Institute (RBI) in Zagreb (HR) on 4H-SiC Schottky diodes [1]. Its electronic features and radiation hardness, combined with a well established process to fabricate nuclear detectors, make 4H-SiC an ideal material to perform such IBIC experiments. Moreover, the exhaustive electronic and ion/x-ray beam analyses performed in the last years on these devices provide well assessed input parameters for the calculation of CCE profiles and maps at the micrometer scale [4].

In the first of these experiments [1], the IBIC characterization of a Schottky diode array was carried out by mapping of the CCE at different bias voltages and led to the accurate definition of the active region underneath the sensitive electrode surrounded by a grounded guard ring.

Here we report on IBIC measurements on coplanar interdigitated electrodes with different sensitive electrode configurations and carried out at two different proton energies. The experimental results are compared with simulations for a deeper comprehension of the physical mechanism of the induced charge collection in multielectrode structures.

*Experimental*

The sample under test consists in a Schottky diode fabricated by Alenia Marconi on an n-type (40 μm thick, donor concentration: $N_D \approx 10^{15}$ cm$^{-3}$,) epitaxial layer produced by CREE Research company [5].

The back Ti-Pt-Au ohmic contact was deposited on the C face of the n+ substrate. The front Schottky contact consists of a pair of Ni$_2$Si/Au interdigitated finger combs with finger sizes of 50 μm in width and 700 μm in length. The gap between the fingers on opposing combs is 20 μm; combs are separated from the guard ring (50 μm in width) by a gap of 68 μm. A picture of the device is reported in Fig. 1.

The rectifying junctions exhibit 1.56 eV Schottky barrier height and reverse current lower than 0.1 nA at 100 V reverse bias voltage.

The bias voltage is applied at the back ohmic electrode with respect to the ground, which is connected to the guard ring and to the comb G. The connection to the charge sensitive preamplifier defines the sensitive electrode. In configuration A the preamplifier is connected to the back electrode and the comb S is grounded (Fig. 2a); in configuration B, it is connected to the comb S (Fig. 2b). The surface plots in Figs. 2a,b indicate the electric potential distribution when a bias voltage of 10 V is applied at the back electrode in configurations A and B, respectively.

The IBIC measurements were carried out at the ion microbeam facility of the RBI using 0.9 MeV and 1.5 MeV proton beams focused to a spot size of less than 1 μm and raster scanned onto the device surface. The ion currents were maintained below 0.1 fA in order to avoid pileup effects and the measurements were performed at room temperature in high vacuum conditions (pressure below $10^{-3}$ Pa).

The measurement of the charge collection efficiency (CCE) was carried out by evaluating the energy deposited by the ion only in the active region (i.e. subtracting the energy lost in the metal contact and in the underlying dead layer [6]) and normalizing the IBIC signal from the SiC diode to the signal generated by a reference silicon surface barrier detector, for which a complete charge collection is assumed.

CCE mapping was calculated with the mathematical approach described elsewhere [1] by using the finite element software package FEMLAB to solve the coupled Poisson's and continuity equations, the relevant weighted potential distribution and the adjoint equations [7]; the relevant physical parameters were extracted from [8]. The IBIC response was calculated by combining the FEMLAB output with the energy loss profiles, as calculated by the SRIM Monte Carlo code [9].

### *Results and discussion*

Figs. 3a,b show the IBIC maps at bias voltage 10 V for the device in configuration A and B, respectively, using 0.9 MeV focused ion beams. In configuration A, the back contact is the sensitive electrode and the IBIC map exhibits both interdigitated comb structures. In configuration B, the charge preamplifier is connected to one of the two frontal electrodes, whereas the other is grounded; the relevant IBIC map exhibits only one comb structure.

To interpret the data, we report in Figs. 4a,b the two dimensional Gunn's weighting potential (GWP) maps evaluated at the same bias voltage in the two configurations.

The GWP is defined as the derivative of the actual electric potential ψ vs. the voltage V applied at the sensitive electrodes ($GWP = \left.\frac{\partial \psi}{\partial V}\right|$) when the potentials of all the other conductors in the system are held constant. According with the basic theory illustrated in [1], the GWP profile defines regions in which the movement of free carriers induces a current at the sensitive electrode.

In configuration A, the GWP coincides with ψ, i.e. Fig. 4a resembles the surface plot in Fig. 2a; this means that any carrier generated in a region where the electric field occurs, induces a charge at the back electrode, whose value is proportional to the actual electric potential. In configuration B, the GWP does not coincide with ψ; it extends underneath the sensitive electrode, whereas it is null underneath the grounded electrode, therefore only carriers moving along force lines converging to the sensitive electrode induce a measurable charge. As a consequence, IBIC signals can be detected only when ions are incident on the sensitive comb electrode, whereas they are almost null when hit the grounded comb electrode.

The analysis of the CCE profiles along the x direction shown in Fig. 5a,b confirm the above remarks, i.e. the IBIC signals are generated only underneath the sensitive electrodes. Further remarks can be made considering that the charge collection underneath the centres of the sensitive electrodes is complete (i.e. CCE=100%) at a bias voltage higher than 70 V for protons of 0.9 MeV. No differences are observable, within our signal resolution, between the CCE maxima measured in the two configurations. The curve shown in Fig. 6 is relevant to 0.9 MeV proton irradiation and it shows the behaviour of the CCE maxima as function of the applied bias voltage. The CCE grows as the bias voltage increases up to the complete charge collection occurring at bias voltages higher than 70 V. The growing part of the curve can be suitably interpreted by the drift diffusion model, which allows a reliable estimation of the diffusion length of the minority carrier (hole in this case) [2]. The fitting procedure provide a value of $L_d$=6 µm, which corresponds to a hole lifetime of 125 ns, if a hole diffusivity of 3 cm$^2$·s$^{-1}$ is assumed. These values are confirmed by applying the fitting procedure to the data obtained using 1.5 MeV protons. In this case the CCE does not saturate since the generation profile induced by ionisation extends beyond the extension of the depletion region at the maximum applied voltage, i.e. the Bragg's curve extinguishes at 21 µm, whereas at 140 V the depletion region extends 12 µm underneath the center of the sensitive electrode and a significant fraction of the generation profile falls within the neutral region (Fig. 7).

~~The above mentioned minority carrier lifetime value, as well as the other transport parameters of 4H-SiC material [8], have been considered as input parameters for calculating CCE maps through the solution of the adjoint continuity equations of holes and electrons.~~

It is worth stressing that the above mentioned carrier lifetime value and the other transport parameters of 4H-SiC [8] are not the result of an iterative fitting procedure, which would be too demanding in terms of computational resources, but they were instead chosen with a "test and trial" approach as the best input parameters for calculating the CCE maps through the solution of the adjoint continuity equations.

Fig. 8 shows the calculated CCE maps in both configurations for bias voltages of 10, 70 and 130 V. The active region of the detector in both configurations extend beyond the regions shown in Fig. 4 (i.e. where the GWP is not null). In fact, a fraction of the minority carriers generated in the neutral region diffuse towards the depletion region and hence contribute to the induced current. Such a contribution to the IBIC signal has a typical exponential profile and extinguishes at about three times the diffusion length of the minority carriers (in our case 18 µm).

As the applied bias voltage increases, the active region deepens and widens, allowing IBIC pulses to be detected even when ions hit the sample in the gap between the electrodes. The overlapping of the active regions is enhanced in configuration A, due to the small gap separating two adjacent electrodes, whereas, in configuration B, the presence of a "passive" comb (i.e. the grounded central electrode if Fig. 2b) nullify the charge collection efficiency at about half of the gap. Therefore, the adherence of the IBIC maps with the geometrical profiles of the electrodes is enhanced at low bias voltages; in configuration B the comb profiles are smoothed because of the overlapping of the active regions underneath the inter-electrode gaps.

Such remarks can be clearly applied to interpret the experimental CCE profiles in Figs. 5a,b and find a further validation by the maps / profiles / graphs shown in Fig. 9 obtained by convoluting the CCE maps with the Bragg's generation profile relevant to 0.9 MeV protons. It is worth noticing the excellent adherence of the theoretical curves to the experimental profiles, as highlighted in Fig. 10.

Similarly, the analysis of the IBIC profiles obtained using 1.5 MeV protons can be easily carried out convoluting the theoretical profiles which can be extracted from the CCE maps illustrated in Fig. 8, with the Bragg's generation profile shown in Fig. 7.

The theoretical and experimental profiles relevant to 1.5 MeV are shown in Fig. 11. ~~The good adherence of the theoretical profile with the experimental one can be appreciated in Fig. 10 for a bias voltage of 10 V.~~ Finally, it is worth noticing that the profiles relevant to 1.5 MeV protons are more rounded than those obtained with 0.9 MeV protons. This effect has the same interpretation as for the smoothing of the CCE maps at decreasing bias voltages ~~increases~~ (see Fig. 8): when the generation profile extends beyond the active region the diffusion transport process plays a more significant role in the generation of the IBIC signal.

**Conclusions**

In this paper we report on the IBIC analysis of a Schottky diode with coplanar interdigitated electrodes. The experimental results have been interpreted on the basis of a theoretical approach based on the Shockley-Ramo-Gunn theorem described in previous works. The theoretical approach proved to be able to give valuable insight on the charge collection process and allows an exhaustive characterization of the device. Besides providing a robust validation of the theoretical framework for mapping charge pulses generated within the whole detector volume, this work provides direct evidence of the important role played by diffusion in the charge collection process and in the definition of the active areas in pixellated electrode structures.

**Acknowledgements**

This work has been carried out and financially supported in the framework of the INFN experiment "DANTE". The work of P. Olivero was supported by a "visiting scientist" scholarship funded by the Ruđer Bošković Institute, which is gratefully acknowledged.

*Figure Captions*

Fig. 1   Photograph of the planar interdigitated Schottky electrodes. .

Fig. 2   Cross section scheme of the device with the electrical connections; configuration A: the sensitive electrode is the back contact; configuration B: the sensitive

electrode is the front comb S. The surface plot represents the electric potential distribution at an applied bias voltage of 10 V.

Fig. 3    IBIC map of the device in A (Fig. 2a) and B (Fig. 2b) configuration; bias voltage = 10 V, proton energy=900 keV

Fig. 4    Gunn's weighting potential maps relevant to configurations A (4a) and B ( 4b) at bias voltage V=10 V.Fig. 5    CCE vs. applied bias voltage curves relevant to ions of 0.9 MeV (curve A)  and 1.5 MeV (curve B) hit the center of the top electrode S. The continuous lines are the fitting curves calculated by the drift-diffusion method.

Fig. 5: experimental CCE profiles along the x direction in the electrode configurations A and B using 0.9 MeV protons at two bias voltages. The vertical bars at x=-45,-25,+25,+45 $\mu$m indicate the border of the electrodes shown in Fig. 2.

Fig. 6    CCE vs. applied bias voltage curves relevant to ions of 0.9 MeV (curve A)  and 1.5 MeV (curve B) hit the center of the top electrode S in configuration A. The continuous lines are the fitting curves calculated by the drift-diffusion method.

Fig. 7: Ionization profiles (continuous curves; vertical scale on the left) for 0.9 and 1.5 MeV protons in 4H-SiC. The dashed line indicate the extension of the depletion layer region as function of the applied bias voltage (vertical scale on the right).

Fig. 8    CCE maps relevant to configurations A and B at bias voltages (from top to bottom) V=10 V, 70 V  and  V=130 V.

Fig. 9: theoretical CCE profiles along the x direction in the electrode configurations A and B using 0.9 MeV protons at different bias voltages. The vertical bars at x=-45,-25,+25,+45 $\mu$m indicate the border of the electrodes shown in Fig. 2.

Fig. 10: Comparison of the theoretical and experimental CCE profiles for the two configurations and with ions of 900 and 1500 keV. The applied bias voltage is 10 V.

Fig. 11: experimental and theoretical CCE profiles along the x direction in the electrode configurations A 0.9 MeV protons at different bias voltages. The vertical bars at x=-45,-25,+25,+45 μm indicate the border of the electrodes shown in Fig. 2.

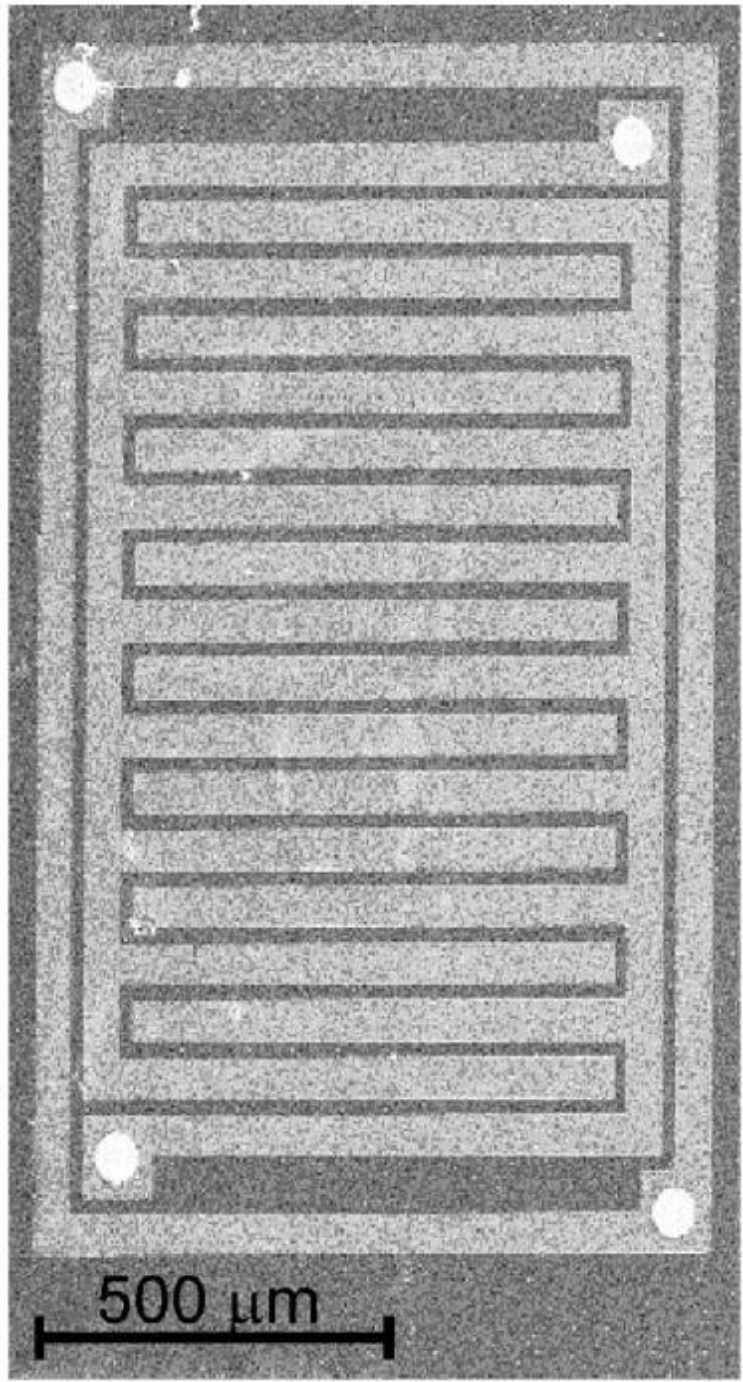

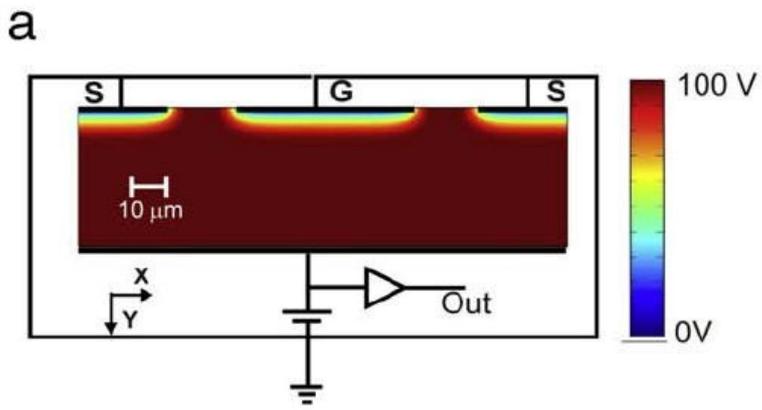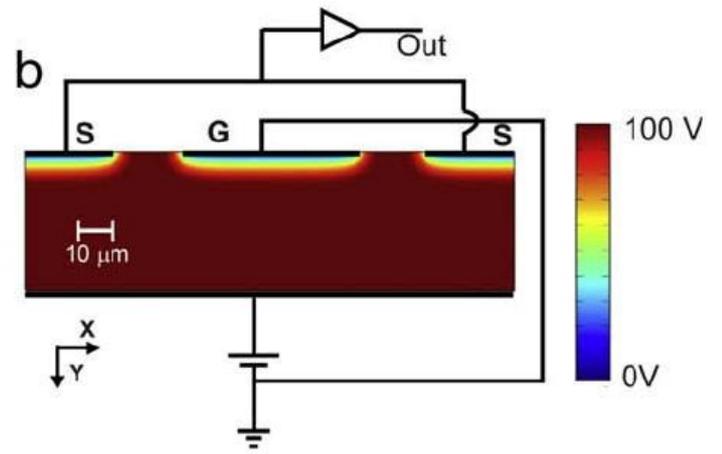

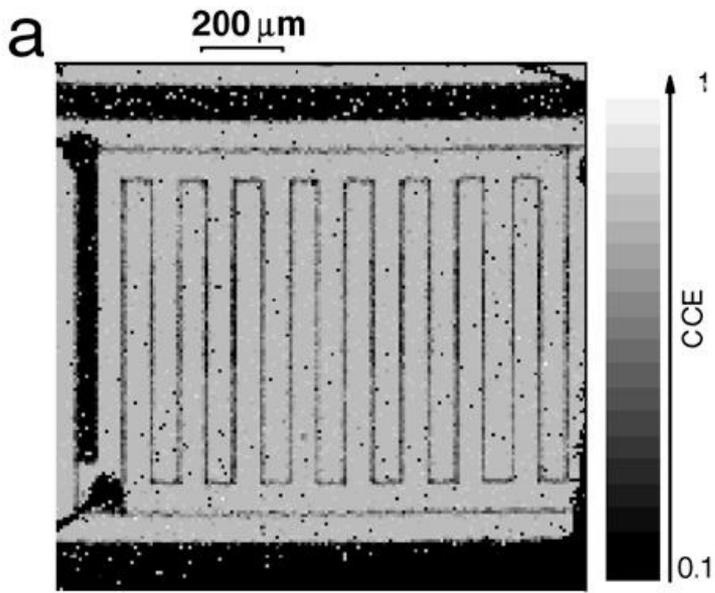 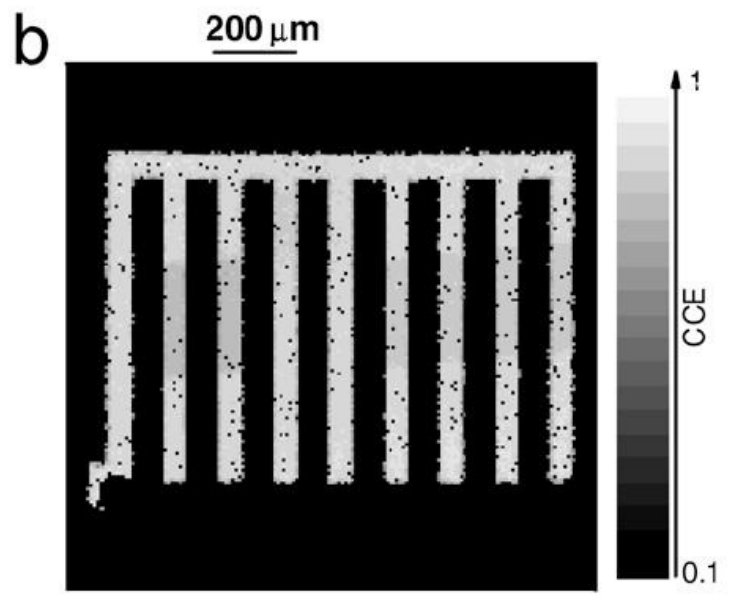

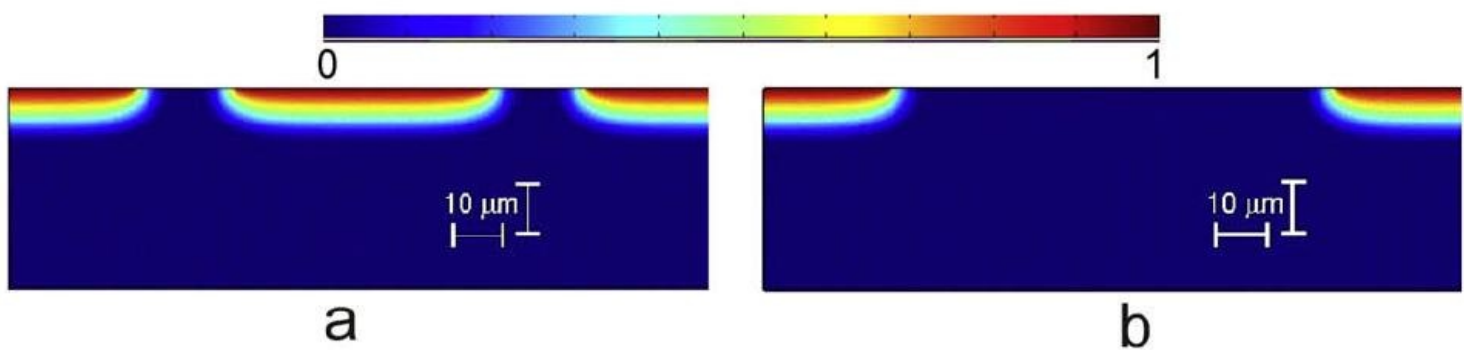

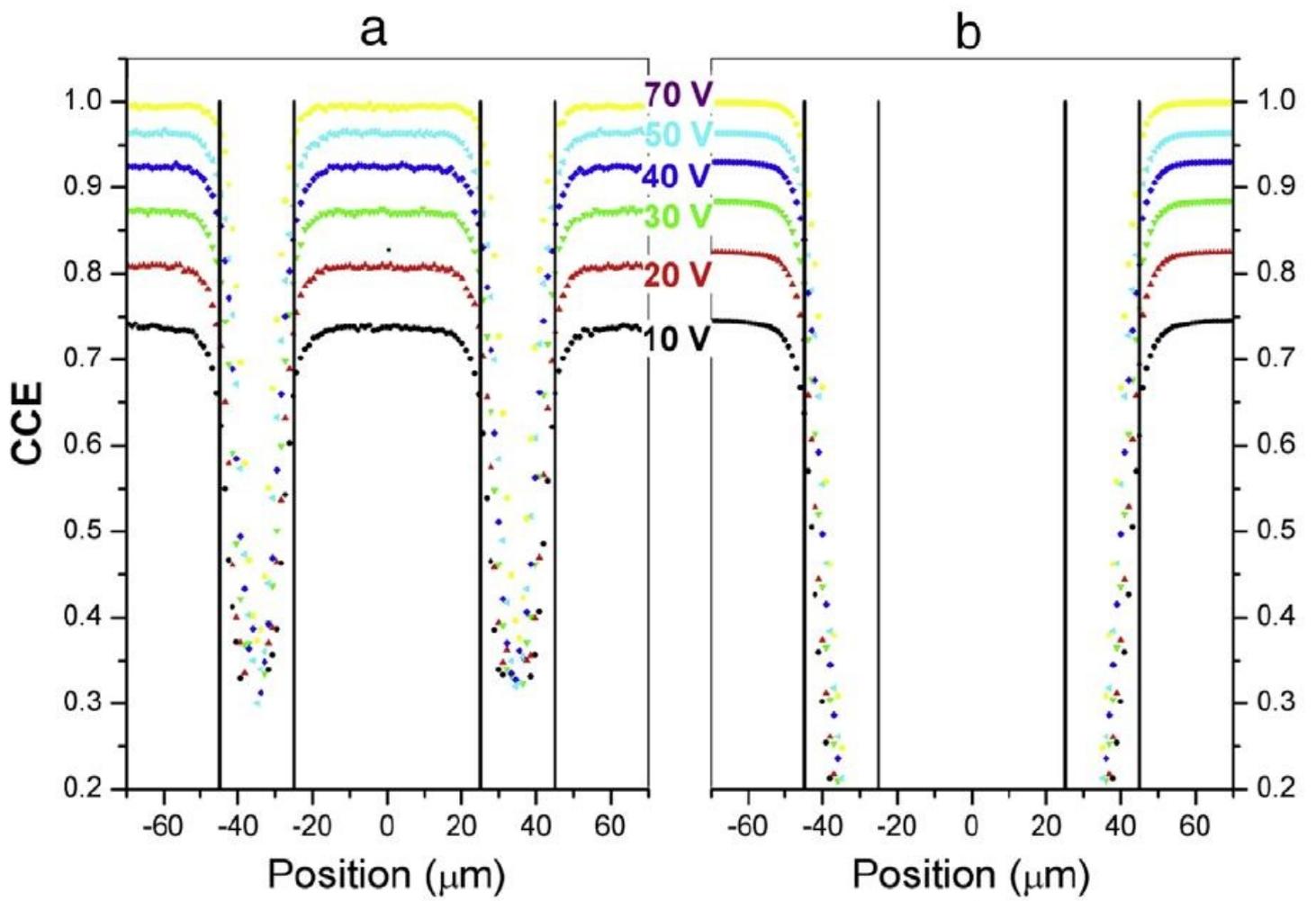

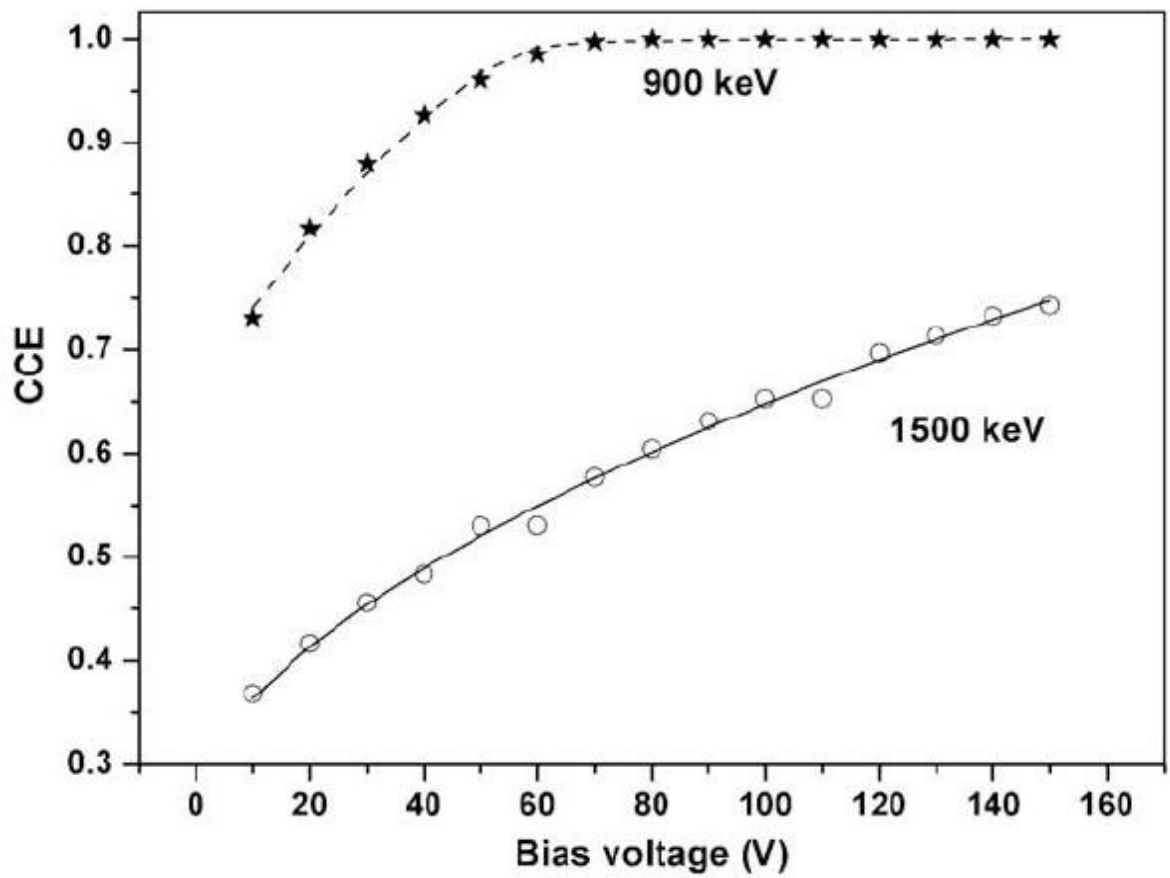

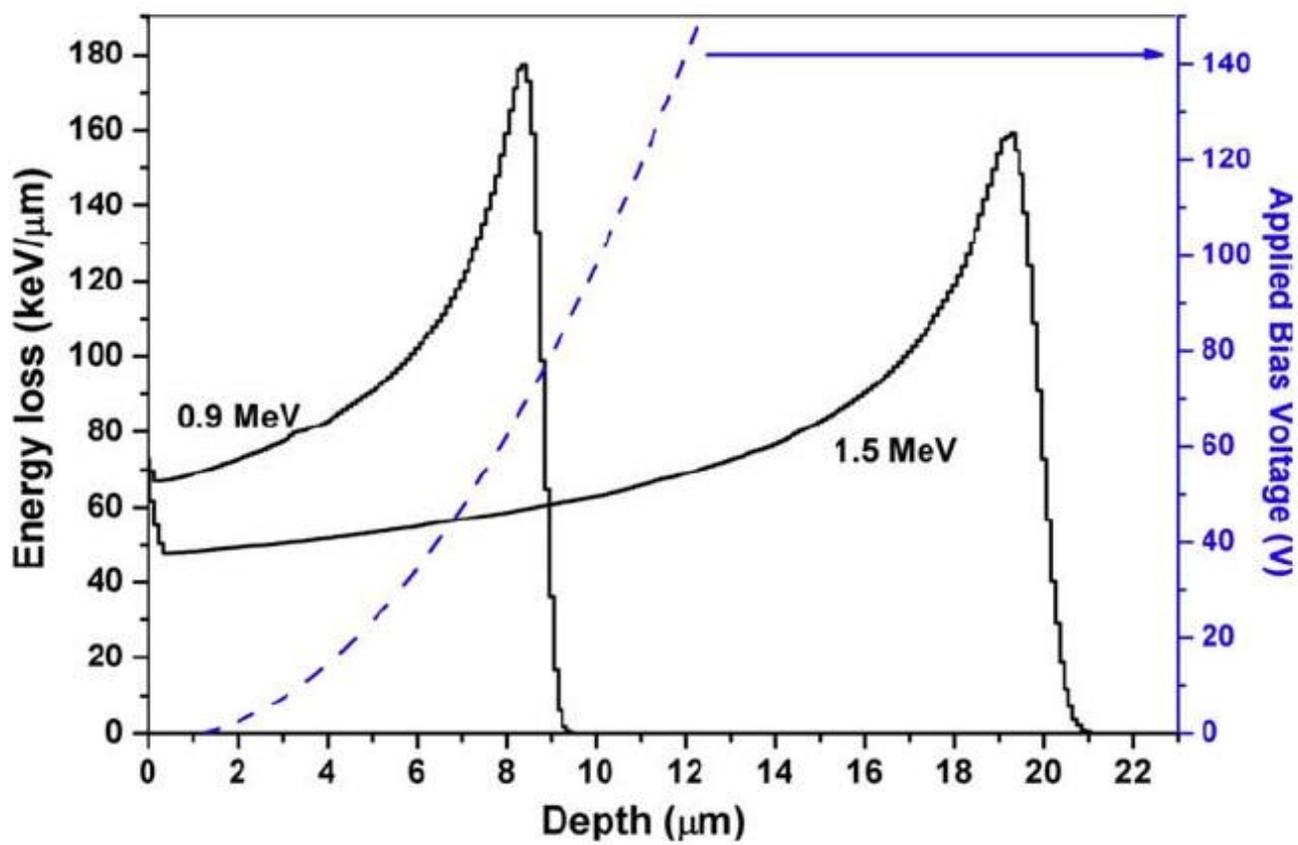

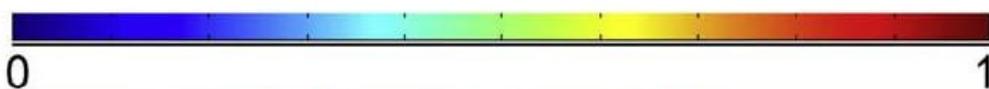
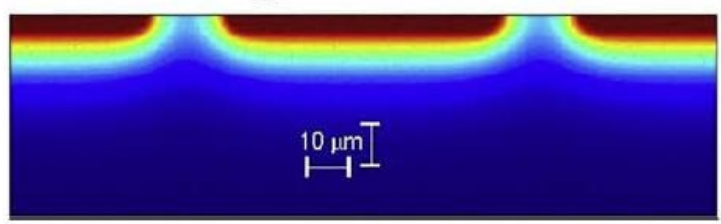
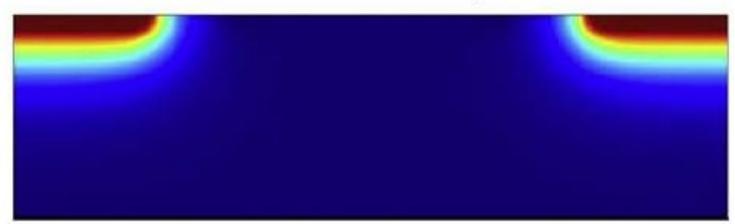
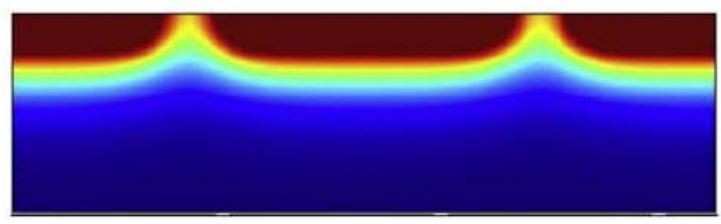
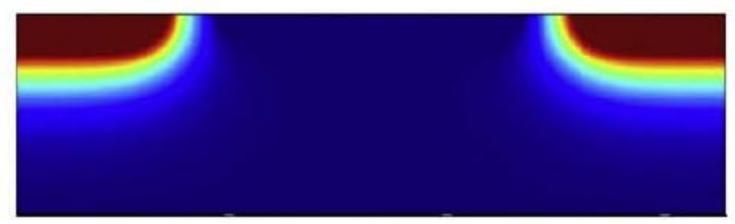
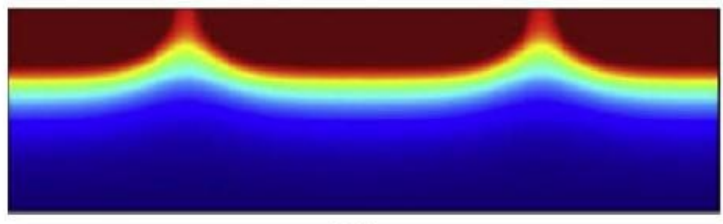
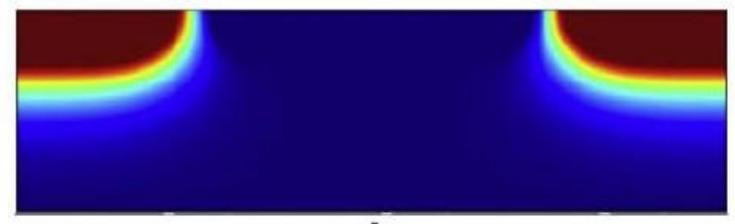

a b

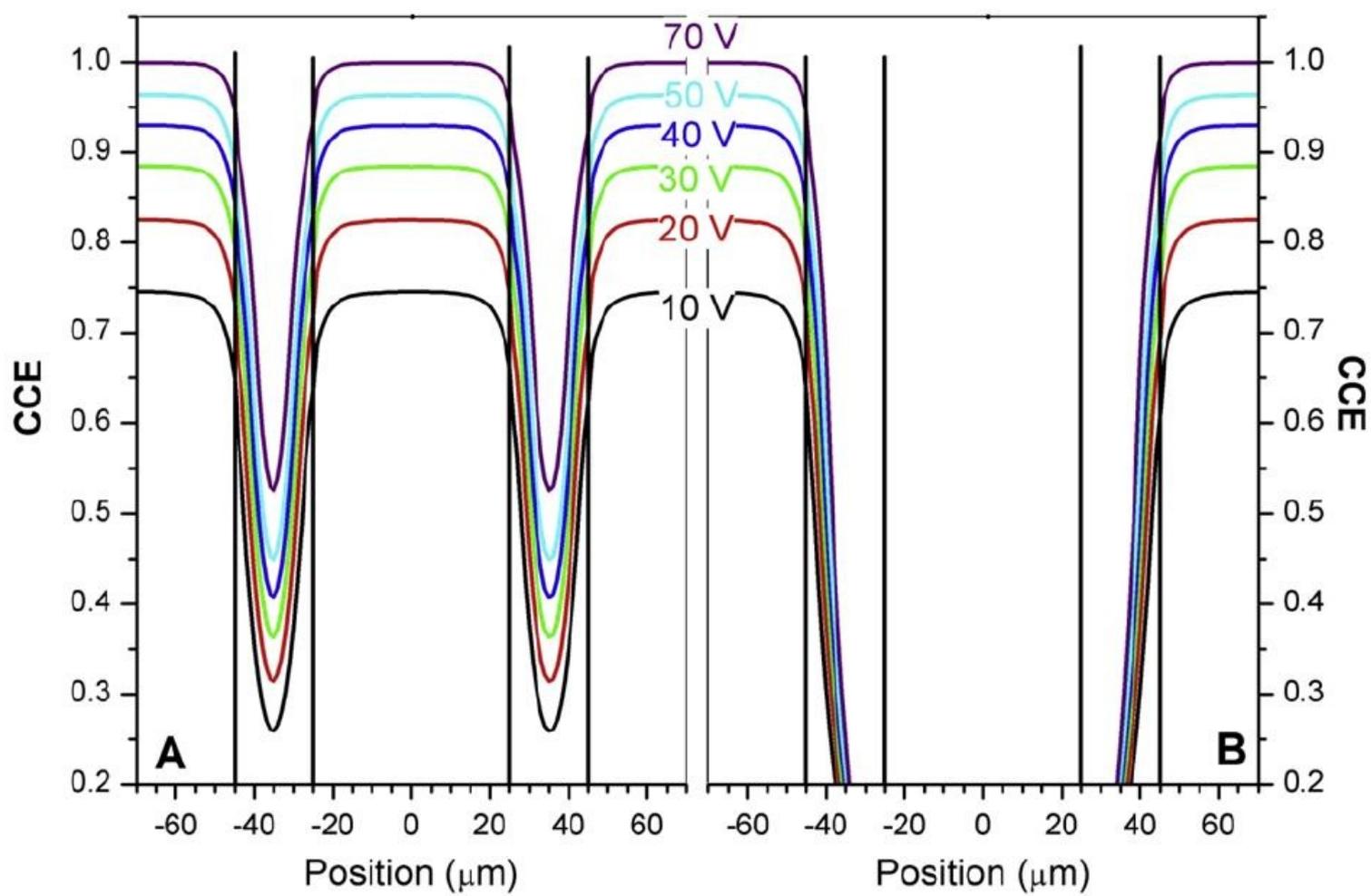

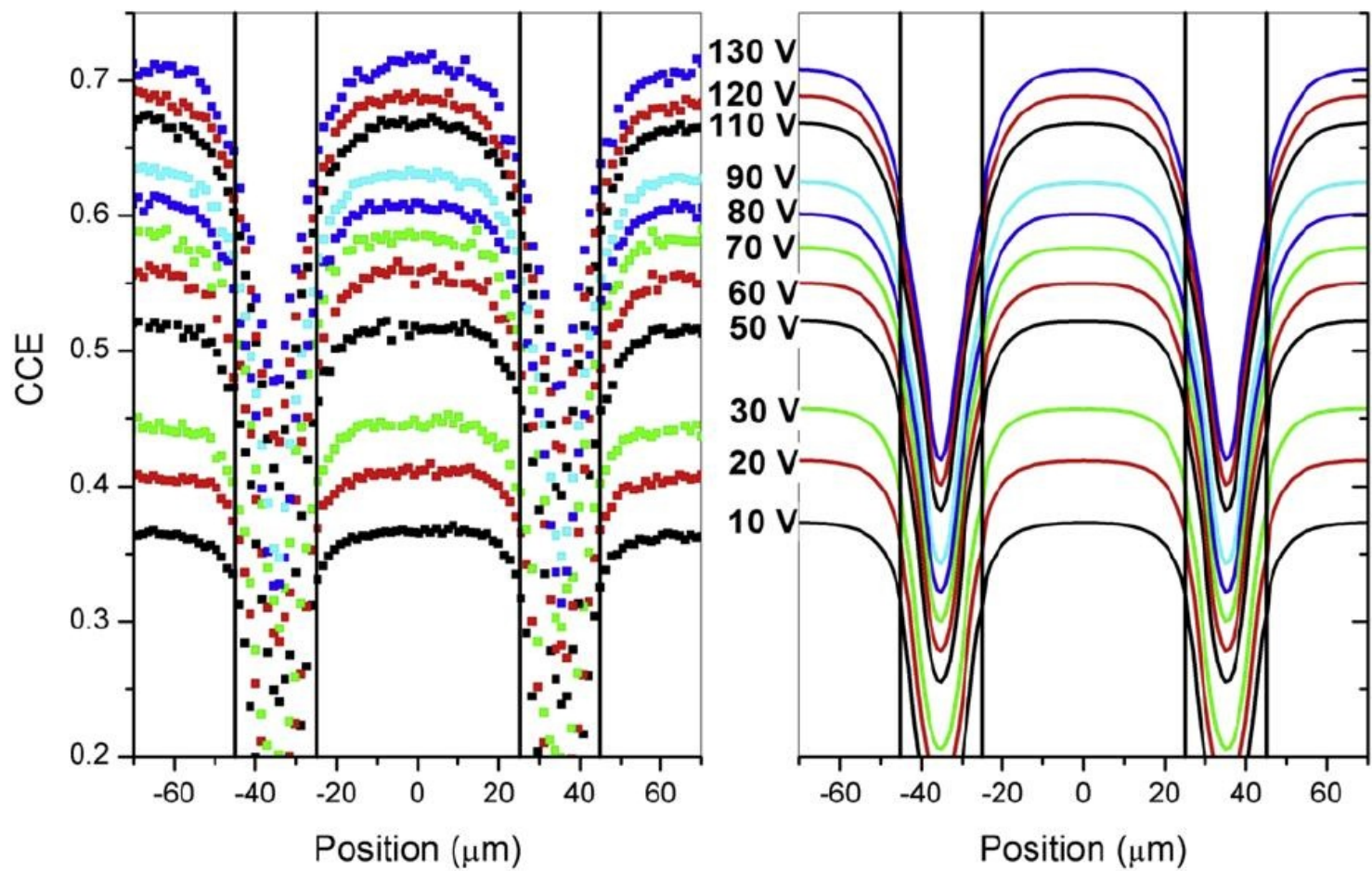

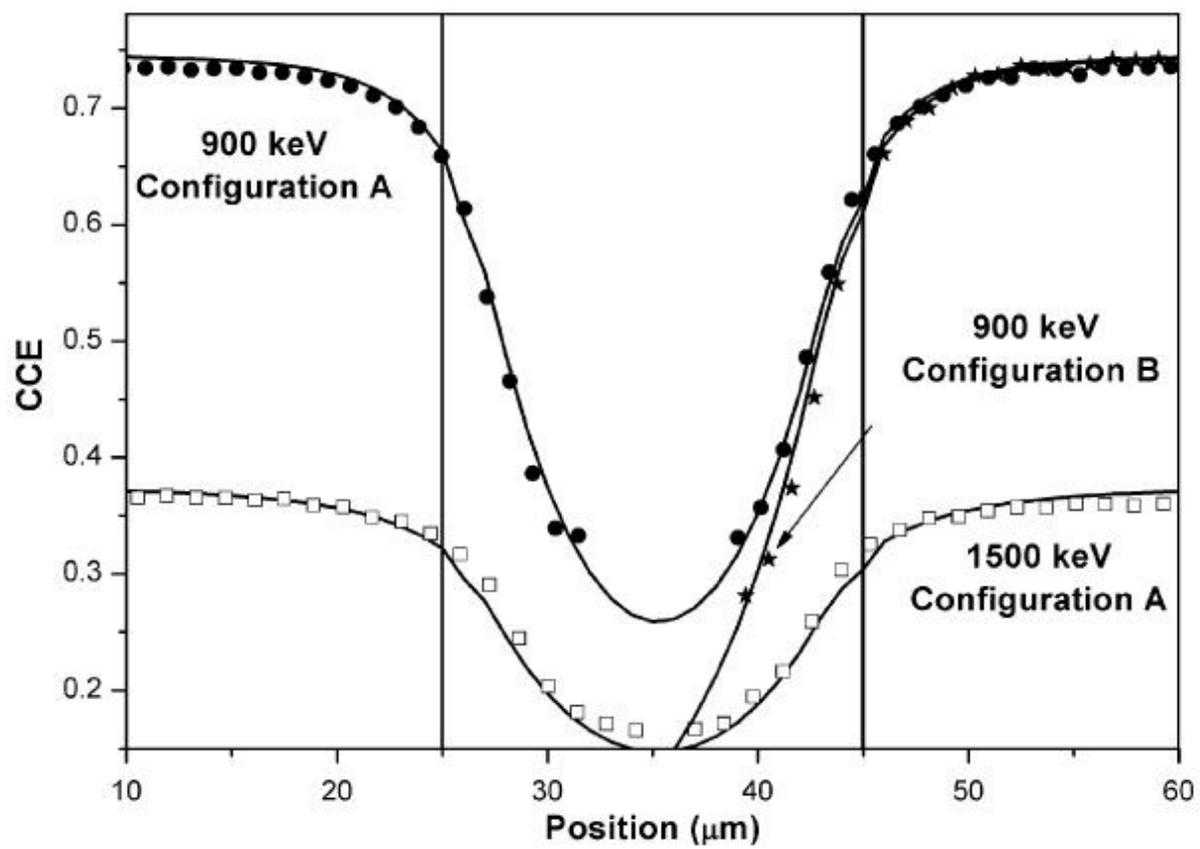